\documentclass[
reprint,
superscriptaddress,
showpacs,preprintnumbers,
amsmath,amssymb,
aps,
]{revtex4-1}

\usepackage{color}
\usepackage{graphicx}
\usepackage{dcolumn}
\usepackage{bm}
\usepackage{txfonts}
\usepackage{tikz}
\usepackage{pgffor}
\usepackage{verbatim}
\usepackage{pstricks}
\usepackage{pst-3dplot}
\usepackage[colorlinks, breaklinks=true, linkcolor=blue, citecolor=blue, linktocpage=true]{hyperref}

\begin{document}


\title{Gain-driven magnon-polariton dynamics in the ultrastrong coupling regime:\\ 
Effective circuit approach for coherence versus nonlinearity
}

\author{Ryunosuke Suzuki}
 \affiliation{Department of Applied Physics, Graduate School of Engineering, Tohoku University, Sendai, Miyagi 980-8579, Japan}
\author{Takahiro Chiba}
 \affiliation{Department of Information Science and Technology, Graduate School of Science and Engineering, Yamagata University, Yonezawa, Yamagata 992-8510, Japan}
 \affiliation{Department of Applied Physics, Graduate School of Engineering, Tohoku University, Sendai, Miyagi 980-8579, Japan}
\author{Hiroaki Matsueda}
 \affiliation{Department of Applied Physics, Graduate School of Engineering, Tohoku University, Sendai, Miyagi 980-8579, Japan}
 \affiliation{Center for Science and Innovation in Spintronics, Tohoku University, Sendai 980-8577, Japan}


\date{\today}

\begin{abstract}
We theoretically study the dynamics of gain-driven magnon-polaritons (MPs), which characterizes auto-oscillation of MPs, across the strong coupling (SC) and ultrastrong coupling (USC) regimes. Taking into account the magnon dynamics via the magnetic flux, we present an effective circuit model of gain-driven MPs, which allows to manipulate the coupling strength of MPs by tuning the size of a ferromagnet and incorporates the self-Kerr nonlinearity of magnons due to the shape magnetic anisotropy.
In the SC regime, we find that the self-Kerr nonlinearity generates a frequency shift and reduces the coherent magnon-photon coupling. 
In contrast, in the USC regime, we find that the coherent magnon-photon coupling not only overcomes the self-Kerr nonlinearity but also effectively couples to gain via the imaginary part of complex eigenfrequencies, resulting in magnon-like auto-oscillations. 
Subsequently, the USC enables one to widely tune the auto-oscillation frequency by means of an external magnetic field. 
These findings indicate that there is a trade-off relation between the coupling strength of MPs and the self-Kerr nonlinearity of magnons.
This work is attributed to understanding of the interplay between gain-loss and USC in nonlinear polariton dynamics, offering a novel principle for frequency tunable maser-like devices based on gain-driven MPs.
\end{abstract}

\maketitle

\section{Introduction}

Cavity magnonics is one of the prominent platforms for hybrid-quasiparticle physics, wherein the key excitation is magnon-polariton (MP), that is, a hybridized quasiparticle consisted of electromagnetic fields (photons) in a cavity and collective spin excitations (magnons) in a magnet \cite{Harder21,Rameshti22,Yuan22}. MPs exhibit rich dynamics and enhanced controllability, which have inspired the development of novel quantum devices such as quantum detectors \cite{Quirion20}, quantum memories \cite{Zhang15,Shen21}, and maser-like devices \cite{Hou21,Yao23}.

As an important point for practical applications, stronger coherent magnon-photon coupling enables faster and more efficient control of quantum states. However, the achievable coupling strength is fundamentally limited by the large mode volume of microwave cavities relative to the small size of typical magnets, resulting in coupling strengths ($g$) that are much smaller than the natural angular frequency ($\omega_{\rm r}$) of either the photon mode or the magnon mode. Therefore, realizing the so-called ultrastrong coupling (USC) regime \cite{Diaz19,Kockum19}---characterized by a coupling-to-frequency ratio $g/\omega_{\rm r} \gtrsim 0.1$---is highly desired, which opens up a new frontier in hybrid light-matter interactions \cite{Baydin25}. The USC regime is typically characterized by the vacuum Bloch-Siegert shift \cite{Diaz10,Li18} on polariton eigenfrequencies, which emerges in the framework beyond the rotating-wave approximation (RWA) and associates with the excitation number nonconserving process in the quantum level. Very recently, USC in MP systems has been achieved experimentally in structure-based platforms with various shapes of magnets, such as supperconductor/ferromagnet nanostructures \cite{Golovchanskiy21SciAdv,Golovchanskiy21PRAP,Ghirri23} , slab geometries \cite{Bourcin23}, and magnetic metamaterials \cite{Mita25PRAP,Mita25arXiv}, thereby overcoming the limitation of conventional cavity-based MP systems.

Recently, gain-embedded cavity magnonics has emerged as a platform to engineer a gain-driven MP \cite{Yao23,Zhang25}, which displays auto-oscillation of polaritons and self-selection of a sustained oscillation mode. In the viewpoint of practical applications, the gain-driven MP possesses advantageous features, involving high coherence and high power output \cite{Yao23,Kim24,Gui24}, injection locking \cite{Yao23,Lu24}, microwave amplification \cite{Yao23,Lu24}, frequency-comb generation \cite{Gui25}, and meterscale strong coupling \cite{Rao23}.
Despite its potential, the theoretical treatment of the gain-driven MPs remains incomplete. Most existing models are based on a two-mode version of the Tavis-Cummings model in the framework of RWA, which assumes the strong coupling (SC) regime and treats the dynamics of MPs as a linear response. Consequently, intrinsic nonlinear effects, especially magnonic self-Kerr nonlinearity arising from the magnetic anisotropy in magnets \cite{Zheng23,Elyasi20,Tatsumi25}, are often neglected. 
Experimentally, the self-Kerr nonlinearity arises in nonlinear systems, including cavity magnonic platforms \cite{Wang16,Wang18}, when subjected to sufficiently strong external driving fields. Then, the self-Kerr nonlinearity typically manifests as a power-dependent shift in the resonance frequency.
As a consequence, the conventional Tavis-Cummings-type models fail to fully capture dynamics of gain-driven MPs, particularly in regimes where the self-Kerr nonlinearity and the USC play a critical role. 
A more comprehensive description of the gain-driven MP thus requires incorporating the nonlinear response of magnetization dynamics into the theoretical framework, especially when aiming to explore operation regimes beyond the SC regime.

In this article, we theoretically demonstrate that the role of magnonic self-Kerr nonlinearity in gain-driven MPs varies significantly across coupling regimes. To this end, we consider a gain-embedded cavity magnonics system in which a gain is described by a negative differential resistance and MPs are described by an inductor with inserted ferromagnet. This effective circuit model incorporates the self-Kerr nonlinearity of magnons due to the shape magnetic anisotropy of the ferromagnet and allows to manipulate the coupling strength of MPs by tuning the size of the ferromagnet.
In the SC regime, we show that the self-Kerr nonlinearity dominates the dynamics of the gain-driven MP, resulting in a nonlinear frequency shift and a suppression of the coherent magnon-photon coupling. Consequently, the gain-driven MP mostly reduces to an original auto-oscillation of the photon-mode. In contrast, in the USC regime, the self-Kerr nonlinearity is effectively suppressed by the coherent magnon-photon coupling, leading to stable auto-oscillations of the MP. Moreover, we show that the coherent magnon-photon coupling effectively couples to gain via the imaginary part of complex eigenfrequencies, enabling to control the auto-oscillation frequency by means of an external bias magnetic field. We also discuss a trade-off relation between the coupling strength of MPs and the self-Kerr nonlinearity of magnons as well as the interplay between gain-loss and USC in nonlinear polariton dynamics.

This article is organized as follows: In Sec.~\ref{Model}, we introduce an effective circuit model which describes gain-driven MPs with magnonic self-Kerr nonlinearity.
In Sec.~\ref{Kerr nonlinearity and gain}, based on the circuit model, we discuss effects of the self-Kerr nonlinearity and gain on the eigenmode of a gain-driven MP.
In Sec.~\ref{Numerical demonstration}, in order to demonstrate the nonlinear and gain effects discussed in Sec.~\ref{Kerr nonlinearity and gain}, we numerically investigate the transient response and auto-oscillation of gain-driven MPs in the SC and USC regimes.
Finally, in Sec.~\ref{Conclusion} we summarize our results and discuss the scope of our study.

\begin{figure}[ptb]
\begin{centering}
\includegraphics[width=0.45\textwidth,angle=0]{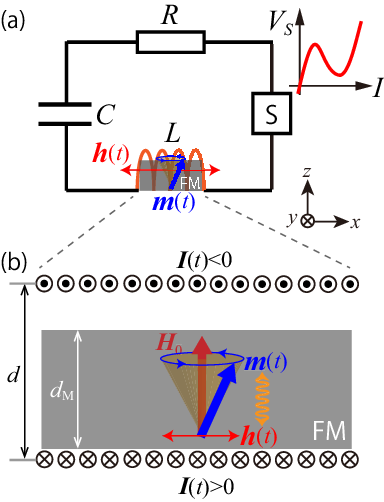} 
\par\end{centering}
\caption{
(a) Effective circuit model of gain-embedded cavity magnonics system in which ${\bm h}(t)$ is a microwave magnetic field (photon) in an inductor and $\bm{m}(t)$ is nonlinear magnetization dynamics (magnon) of a ferromagnet (FM) with the shape magnetic anisotropy. 
$S$ represents an ``S-type'' negative resistance element biased at the operating point. 
(b) Schematic image of the dynamics of gain-driven magnon-polaritons. Here, the cross-section of the inductor is assumed to be rectangular, which is characterized by $d$ and $d_{\rm M}$ being thickness of the inductor and ferromagnet, respectively, as well as the common width $w$. 
}
\label{Fig:model}
\end{figure}

\section{Effective circuit model}
\label{Model}

We begin by constructing an effective circuit model of gain-driven MPs. The circuit model consists of a single-mode $LC$ resonator with an inductor involving a ferromagnet and an amplifier circuit with a negative differential resistance component. The schematic image of the circuit is shown in Fig.~\ref{Fig:model}. In the circuit model, MP is expressed by the dynamics of the $LC$ resonator coupled to a ferromagnet  \cite{Bai15,Grigoryan18,Chiba24APL,Chiba24MSJ,Chiba25}. The negative differential resistance component gives a gain into the circuit, which drives auto-oscillation of an electrical current modeled by a van der Pol (vdP) oscillator with a single
limit cycle. The electric current flowing in the inductor generates a microwave magnetic field via Amp\`{e}re's law, which is treated as the photonic degree of freedom. Also, this microwave magnetic field drives precessional motion of the magnetization in the ferromagnet, which corresponds to the magnonic degree of freedom and is governed by the Landau-Lifshitz-Gilbert (LLG) equation, 
\begin{equation}
\label{LLGeq}
    \frac{d\bm{m}}{dt}=-\gamma\bm{m}\times\left(  -\frac{1}{M_{\rm s}}\frac{\delta U_{\rm m}}{\delta\bm{m}}+\mu_{0}\bm{H}(t)\right)+\alpha\bm{m}\times\frac{d\bm{m}}{dt},
\end{equation}
where $\bm{m}=(m_{x}.m_{y},m_{z})$ is the unit vector aligned with the magnetization direction of the ferromagnet with the saturation magnetization $M_{\rm s}$, $\gamma$ is the gyromagnetic ratio, $\mu_{0}$ is the permeability of vacuum, and $\alpha$ is the intrinsic Gilbert damping constant.
Here, the total magnetic energy is given by
\begin{equation}
\label{magneticenergy}
    U_{\rm m}=-\mu_{0}M_{\rm s}\bm{m}\cdot\bm{H}_{0} + \frac{1}{2}\mu_{0}M_{\rm s}^{2}m_{z}^{2},
\end{equation}
where $\bm{H}_0=H_{0}\hat{\bm{z}}$ 
is a static external magnetic field applied normal to the surface of the ferromagnet. Here, we assume that the ferromagnet has the film shape modeled by the second term in Eq.~ \eqref{magneticenergy}. Also, the term $\bm{H}(t)$ in Eq.~\eqref{LLGeq} represents the microwave magnetic field induced by the alternating current $I(t)$ via Amp\`{e}re's law, which is given by the $RLC$-circuit equation (Kirchihhoff's voltage law)
\begin{equation}
\label{Kirchihhoff'svoltagelaw}
    N\frac{d\Phi(t)}{dt}+RI(t)+\frac{Q(t)}{C}+V_{S}(I)=0,
\end{equation}
where $Q(t)$ is the accumulated charge on a capacitor with an electrostatic capacitance $C$, $R$ represents a resistance in the $RLC$-circuit. Here, $\Phi(t)$ is the magnetic flux inside the inductor, which is given by 
\begin{equation}
    N\Phi(t)=LI(t)+\frac{Ll}{N}\frac{d_{\rm M}}{d}M_{\rm s}m_x,
\end{equation}
where $d_{\rm M}$ is the film thickness of the ferromagnet and $L=\mu_{0} N^{2}wd/l$ is a self-inductance of the inductor with the length $l$, width $w$, thickness $d$, and a turn number $N$. Note that a parameter $d_{\rm M}/d$ is responsible for tuning the total spin number of the ferromagnet.
Also, $V_{S}(I)$ is an electromotive force of a negative resistance component labeled $S$ in Fig.~\ref{Fig:model}~(a), which is expressed as $V_{S}(I)=-R_a(I(t)-I_{\rm B})+r_b(I(t)-I_{\rm B})^{3}$, where $R_a$ is a gain (in units of $\Omega$), $r_b$ is a nonlinear resistance (in units of ${\rm \Omega A^{-2}}$), and $I_{\rm B}$ is a static bias current.  
To ensure dimensional consistency of the physical quantities, we introduce the effective magnetic field along the $x$-axis, $h_x(t) = NI(t)/l/M_{\rm s}$, which yields the vdP equation instead of Eq.~\eqref{Kirchihhoff'svoltagelaw}
\begin{equation}
\label{vdPeq}
    \frac{d^{2}h_x}{dt^{2}}+2\omega_{\rm c}\left(  \beta-\beta_{a}+\beta_{b}h_x^{2}\right)\frac{dh_x}{dt}+\omega_{\rm c}^2h_x=-\frac{d_{\rm M}}{d}\frac{d^{2}m_x}{dt^{2}},
\end{equation}
where $\omega_{\rm c} = 1/\sqrt{LC}$ is the resonance angular frequency of photon, $\beta=R\sqrt{C/L}/2$ is an effective circuit damping, $\beta_{a}=R_a\sqrt{C/L}/2$ is an effective gain, and $\beta_{b}=r_b(M_{\rm s}l/N)^2\sqrt{C/L}/2$ is an effective nonlinear damping. By combining and solving Eqs.~\eqref{LLGeq} and \eqref{vdPeq}, the dynamics of the gain-driven MP can be obtained. As we will see in Sec.~\ref{Numerical demonstration}, this effective circuit model provides a comprehensive framework that captures the self-Kerr nonlinearity, gain, and the USC in the MP system.

Throughout this article, the circuit parameters used in the calculations are as follows:  
$R = 1.0~\mathrm{\Omega}$,  
$C = 1.0~\mathrm{pF}$,  
$l = w = 1.0~\mathrm{mm}$,  
$N = 100$,  
$d = 100~\mathrm{nm}$,  
$R_a = 2.0~\mathrm{\Omega}$, and
$r_b = 20~\mathrm{\Omega A^{-2}}$, which corresponds to 
$\omega_{\rm c}/(2\pi) = 4.49$~GHz,
$\beta = 1.4\times10^{-2}$,
$\beta_a = 2.8\times10^{-2}$, and
$\beta_b = 0.4$.
For simplicity, we also assume $I_{\rm B} = 0$.
The material parameters used in our calculation are those of Yttrium Iron Garnet (YIG), given by
$\gamma = 1.76\times10^{11}~{\rm T^{-1}s^{-1}}$,
$M_{\rm s} = 120~\mathrm{kA/m}$, and
$\alpha = 10^{-4}$ \cite{Schreier15,Bai15}.

\section{Effects of magnonic self-Kerr nonlinearity and gain}
\label{Kerr nonlinearity and gain}

Here, based on the circuit model, we discuss the effects of magnonic self-Kerr nonlinearity and gain on the eigenmode of a gain-driven MP. To this end, we herein deal with approximated LLG equations by assuming a small oscillation of the magnetization.

\subsection{Magnonic self-Kerr nonlinearity}

To facilitate the subsequent analytical treatment and to capture the essential nonlinear behavior of the gain-driven MP, we derive a second-order approximation of Eq.~\eqref{LLGeq} by expanding it in terms of the transverse magnetization (magnon) $m_{\perp} = m_x + i m_y$, under the assumption $m_z \approx 1 - |m_{\perp}|^2 / 2$. This approximation is valid in the regime of relatively small angle precession and enables analytical access to a nonlinear frequency shift and suppression of the coherent magnon-photon coupling.
The resulting approximated form of the LLG equation reads as follows:
\begin{align}
  \label{2ndapproxLLG}
  \frac{dm_{\perp}}{dt}=&-i\tilde{\omega}_{\rm m} m_{\perp}
  +i\gamma\mu_{0}\tilde{M}_{\rm s} h_{x}
  -i\tilde{\alpha}\frac{dm_{\perp}}{dt}-i\frac{\alpha}{2}m_{\perp}\frac{d|m_{\perp}|^{2}}{dt},
\end{align}
where 
\begin{align}
\tilde{\omega}_{\rm m} = 
\begin{cases}
 ~0 & (H_0 < \tilde{M}_{\rm s}) \\
 \gamma\mu_0\left(  H_0 - \tilde{M}_{\rm s}\right) & (H_0 \geq \tilde{M}_{\rm s})
\end{cases}
\label{omegamfilm}
\end{align}
is the resonance angular frequency of magnon including the self-Kerr nonlinearity with
\begin{align}
\tilde{M}_{\rm s} &= M_{\rm s}\left(  1-\frac{|m_{\perp}|^{2}}{2}\right)
,\label{Msm}\\
\tilde{\alpha} &= \alpha\left(  1-\frac{|m_{\perp}|^{2}}{2}\right)
.\label{alpham}
\end{align}
Equation~\eqref{Msm} indicates that a second-order correction ($|m_{\perp}|^2$) modifies the saturation magnetization especially at the steady-state, which manifests the generation of a nonlinear frequency shift and modulation of the coupling strength, as discussed below. Since Eq.~\eqref{2ndapproxLLG} can be rewritten as $dm_{\perp}/dt = -i(\omega_{\rm m} + \gamma \mu_{0}M_{\rm s}/2|m_{\perp}|^2) m_{\perp} \cdots$, where $\omega_{\rm m} = 0~{\rm or}~ \gamma\mu_0(H_0 - M_{\rm s})$ is the eigenmodes of magnon in the linear response, Eq.~\eqref{LLGeq} corresponds to the Duffing-type oscillator in the framework of the second-order approximation \cite{Elyasi20,Tatsumi25,Zhang24van}.
Then, the first term in the right hand side of Eq.~\eqref{2ndapproxLLG} leads to a nonlinear frequency shift proportional to the square amplitude $|m_{\perp}|^2$, from $\omega_{\rm m}$ to $\omega_{\rm m} + \gamma \mu_{0}M_{\rm s}|m_{\perp}|^2/2$, which can be recognized as magnonic version of the Kerr effect \cite{Zhang24van}.  
Also, the second term in the right hand side of Eq.~\eqref{2ndapproxLLG}, corresponding to the coupling to the photon field $h_x$, anticipates that the $m_{\perp}$-nonlinear term modulates the coupling strength via a modified saturation magnetization $\tilde{M}_{\rm s}$ at the steady-state.

In the subsequent analysis, we employ Eq.~\eqref{LLGeq} to perform numerical calculations while we make use of its second-order approximation to obtain analytical insights into the nonlinear oscillation dynamics. It is worth noting that the MP system described by Eqs.~\eqref{vdPeq} and \eqref{2ndapproxLLG} is mathematically equivalent to a fourth-order autonomous dynamical system composing of a Duffing-type oscillator (magnon) coupled to a vdP-type oscillator (photon), which allows the coexistence of several attractors such as fixed points, a limit cycle, and a chaos \cite{Tanekou23}.

\subsection{Complex eigenfrequency including gain}

To derive a excitation condition of the gain-driven MP, we investigate the complex eigenfrequency of MP described by Eqs.~\eqref{LLGeq} and \eqref{vdPeq}. To this end, we decompose ${\bm m} = (m_x,m_y,1)$ by assuming $|m_x|,|m_y| \ll 1$.
Restricting to linear order in $m_x$ and $m_y$, Eq.~\eqref{LLGeq} reduces to a simple oscillator
\begin{align}
\frac{d^2 m_x}{d t^2}
+ 2\alpha\omega_{\rm m}\frac{d m_x}{d t}
+ \omega_{\rm m}^2m_x 
= \gamma\mu_0M_{\rm s}\omega_{\rm m}h_x
,\label{linearLLG}
\end{align}
wherein $\sqrt{1 + \alpha^2}\approx1$ is assumed for $\alpha \ll 1$.
Based on the small oscillation ansatz $h_x(t) = \tilde{h}_xe^{-i\omega t}$ (for linearized \eqref{vdPeq}) and $m_x(t) = \tilde{m}_xe^{-i\omega t}$ \cite{Cheng16}, we obtain the coupled LLG and $RLC$-circuit equations
\begin{align}
\bar{\Omega}
\begin{pmatrix}
\tilde{h}_x \\
\tilde{m}_x
\end{pmatrix}
=
{\bf 0}
\label{LLG-RLC}
\end{align}
with 
\begin{align}
\bar{\Omega} = 
\begin{pmatrix}
\omega^2 + 2i\left(  \beta - \beta_a\right)\omega_{\rm c}\omega - \omega_{\rm c}^2
& (d_{\rm M}/d)\omega^2 \\
\gamma\mu_0M_{\rm s}\omega_{\rm m}
& \omega^2
+ 2i\alpha\omega_{\rm m}\omega 
- \omega_{\rm m}^2
\end{pmatrix}
.\label{OmegaMP}
\end{align}
Hence, the complex eigenfrequencies $\omega_\pm/(2\pi)$ are obtained by solving the determinant of Eq.~\eqref{OmegaMP}.

\begin{figure}[ptb]
\begin{centering}
\includegraphics[width=0.48\textwidth,angle=0]{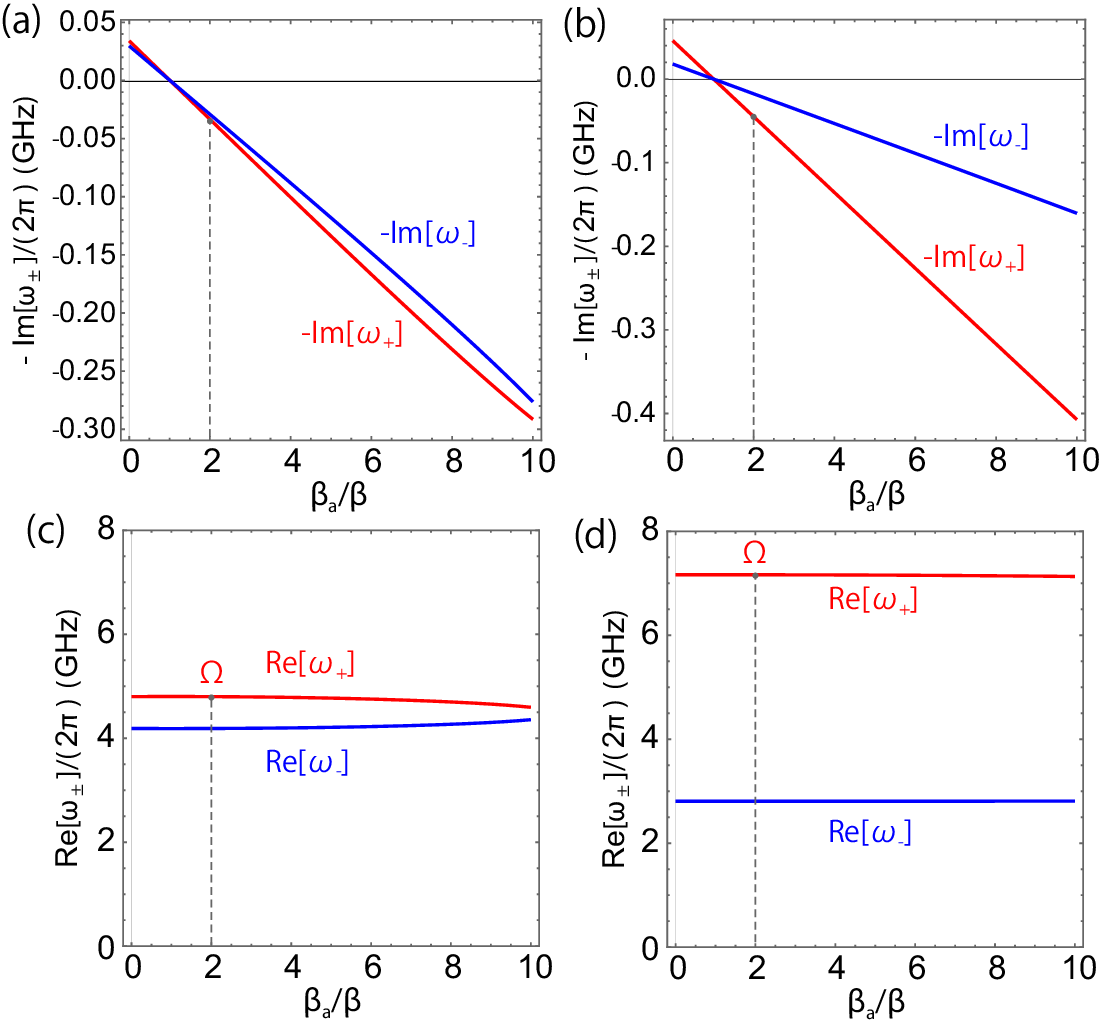} 
\par\end{centering}
\caption{Imaginary part of the complex eigenfrequencies of the linearized gain-driven MP at the original modes crossing point ($\omega_{\rm m} = \omega_{\rm c}$) as a function of a normalized gain ($\beta_a/\beta$). (a) A SC case with $d_{\rm M}/d = 0.02$ and (b) an USC case with $d_{\rm M}/d = 1$.
Real part of the complex eigenfrequencies of the linearized gain-driven MP  at the original modes crossing point as a function of a  normalized gain ($\beta_a/\beta$). (c) A SC case with $d_{\rm M}/d = 0.02$ and (d) an USC case with $d_{\rm M}/d = 1$.
Auto-oscillation ($\Omega$) occurs at the region of $-{\rm Im}[\omega_{\pm}]\leq0$. For $\beta_a/\beta = 2$, ${\rm Re}[\omega_{+}]$ mode is excited by the self-selection.}
\label{Fig:ReImomega}
\end{figure}

For the case of small damping such as $|\beta - \beta_a| \ll 1$ and $\alpha \ll 1$, the real part of the complex eigenfrequency is associated with the coherent magnon-photon coupling. Then, the coupling strength at the original modes crossing point ($\omega_{\rm m} = \omega_{\rm c}$) is defined as
\begin{equation}
g 
\equiv \frac{{\rm Re}\left[\omega_{+}\right] - {\rm Re}\left[\omega_{-}\right]}{2}\Biggl|_{\omega_{\rm m}=\omega_{\rm c}}
\approx \frac{1}{2}\sqrt{\frac{d_{\rm M}}{d}\gamma\mu_0M_{\rm s}\omega_{\rm c}}
.\label{coupling strength}
\end{equation}
On the other hand, the imaginary part of the complex eigenfrequency characterizes the energy dissipation of the system: $-{\rm Im}[\omega_{\pm}] > 0$ corresponds to loss and $-{\rm Im}[\omega_{\pm}] \leq 0$ is gain. For a given external magnetic field, auto-oscillation occurs at the region of $-{\rm Im}[\omega_{\pm}]\leq0$. 
In Figs.~\ref{Fig:ReImomega}~(a) and (b), we plot the imaginary part of the complex eigenfrequencies at the original modes crossing point in SC and USC cases, respectively. Based on Eq.~\eqref{coupling strength}, Fig.~\ref{Fig:ReImomega}~(a) is a strongly coupled case with ${d_{\rm M}/d} = 0.02$ corresponding to $g/\omega_{\rm c} = 0.07$ and Fig.~\ref{Fig:ReImomega}~(b) is an ultrastrongly coupled case with ${d_{\rm M}/d} = 1$ corresponding to $g/\omega_{\rm c} = 0.49$. For a normalized gain $\beta_a/\beta = 2$~(corresponding to $R_a = 2$~$\Omega$ and $R = 1$~$\Omega$), one can see the relation of $-{\rm Im}[\omega_{+}] < -{\rm Im}[\omega_{-}]$ in both cases. Accordingly, $\omega_{+}$-mode that obtains a larger gain is energetically more stable and thereby is excited. This self-selection can be understood by considering the complex mode splitting at the original modes crossing point [see Appendix~\ref{complex mode splitting}]
\begin{equation}
\label{Shift}
\frac{\omega_{+} - \omega_{-}}{2}\biggl|_{\omega_{\rm m}=\omega_{\rm c}} 
= g - \frac{1}{2}g\left(  \alpha + \beta - \beta_{a}\right)i,
\end{equation}
in which a finite imaginary part generates a difference of the lifetime between the $\omega_\pm$-modes, resulting in the self-selection.
Figures~\ref{Fig:ReImomega}~(c) and (d) show the real part of the complex eigenfrequencies at the original modes crossing point as a function of $\beta_a/\beta$. For $\beta_a/\beta = 2$, auto-oscillation frequencies $\Omega/(2\pi)$ are $4.80$~GHz for the SC case and $7.16$~GHz for the USC case, whose values appear to be mostly independent of $\beta_a/\beta$ in both cases. Note that this independence sustains as long as the threshold gain for ${\rm Im}\left[\omega_{\pm}\right]=0$ is relatively small and far form the exceptional point which leads to the level attraction between $\omega_\pm$-modes \cite{Zhang25,Cheng16}.

\section{Numerical demonstration}
\label{Numerical demonstration}

To demonstrate the nonlinear and gain effects discussed in the previous section, we simulate the transient response of gain-driven MPs with the magnonic Kerr nonlinearity. Without loss of generality, we consider the strongly coupled case with ${d_{\rm M}/d} = 0.02$~ (corresponding to $g/\omega_{\rm c} = 0.07$) and the ultrastrongly coupled case with ${d_{\rm M}/d} = 1$~($g/\omega_{\rm c} = 0.49$).
Also, we numerically investigate behaviors of the auto-oscillation frequency of gain-driven MPs with the self-Kerr nonlinearly as a function of an external magnetic field.

\begin{figure*}[ptb]
\begin{centering}
\includegraphics[width=0.99\textwidth,angle=0]{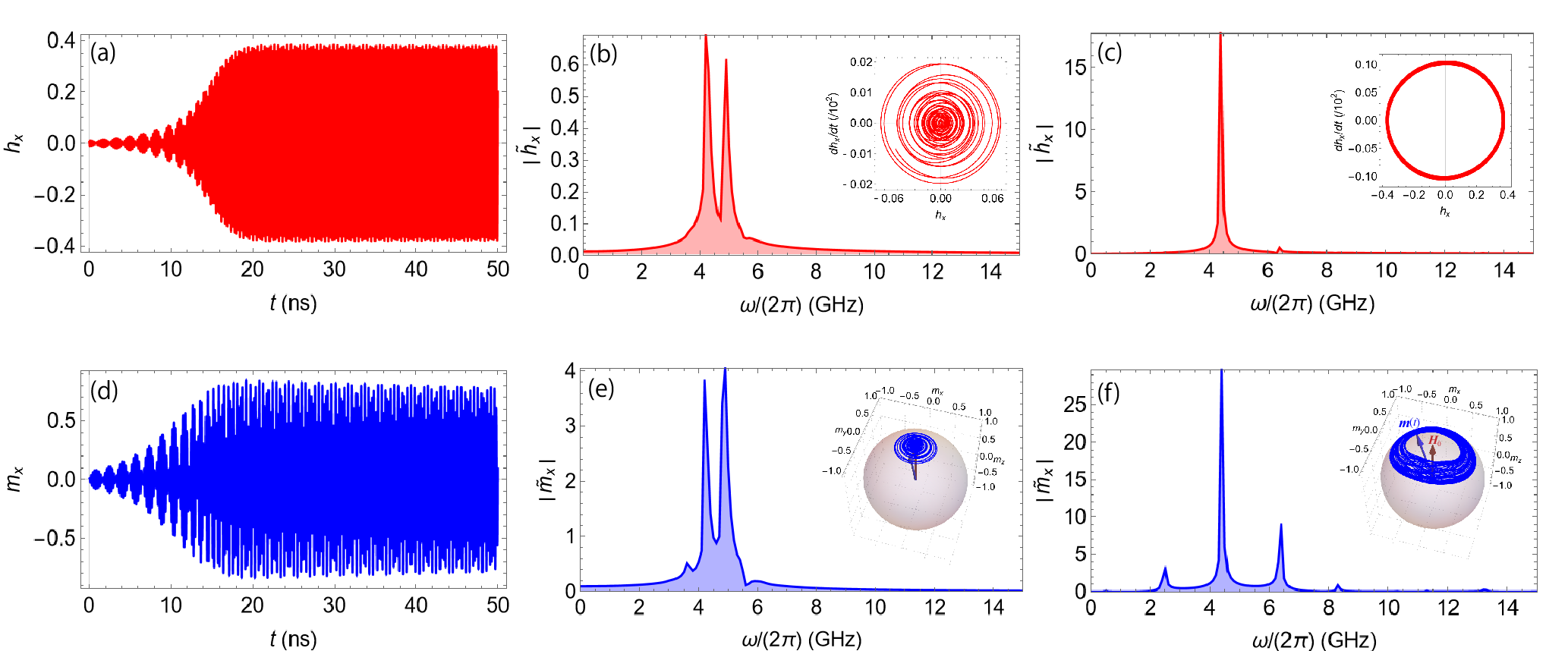} 
\par\end{centering}
\caption{Time evolution and its Fourier spectra of the gain-driven MP in the SC case with $d_{\rm M}/d = 0.02$.
(a) Time evolution of the photon amplitude.
(b) Fourier spectrum of (a) in the time domain \(0 \leq t \leq 10\,\mathrm{ns}\).
(c) Fourier spectrum of (a) in the time domain \(40 \leq t \leq 50\,\mathrm{ns}\).
(d) Time evolution of the magnon amplitude.
(e) Fourier spectrum of (d) in the time domain \(0 \leq t \leq 10\,\mathrm{ns}\).
(f) Fourier spectrum of (d) in the time domain \(40 \leq t \leq 50\,\mathrm{ns}\).
Insets show the corresponding phase space trajectories of the photon and magnon.  
The calculations are performed with the initial condition of $h_x(0) = 0.01$ and $m_x(0) = 0$ at the original modes crossing point ($\omega_{\rm m} = \omega_{\rm c}$).
}
\label{Fig:SCGMP}
\end{figure*}

\begin{figure*}[ptb]
\begin{centering}
\includegraphics[width=0.99\textwidth,angle=0]{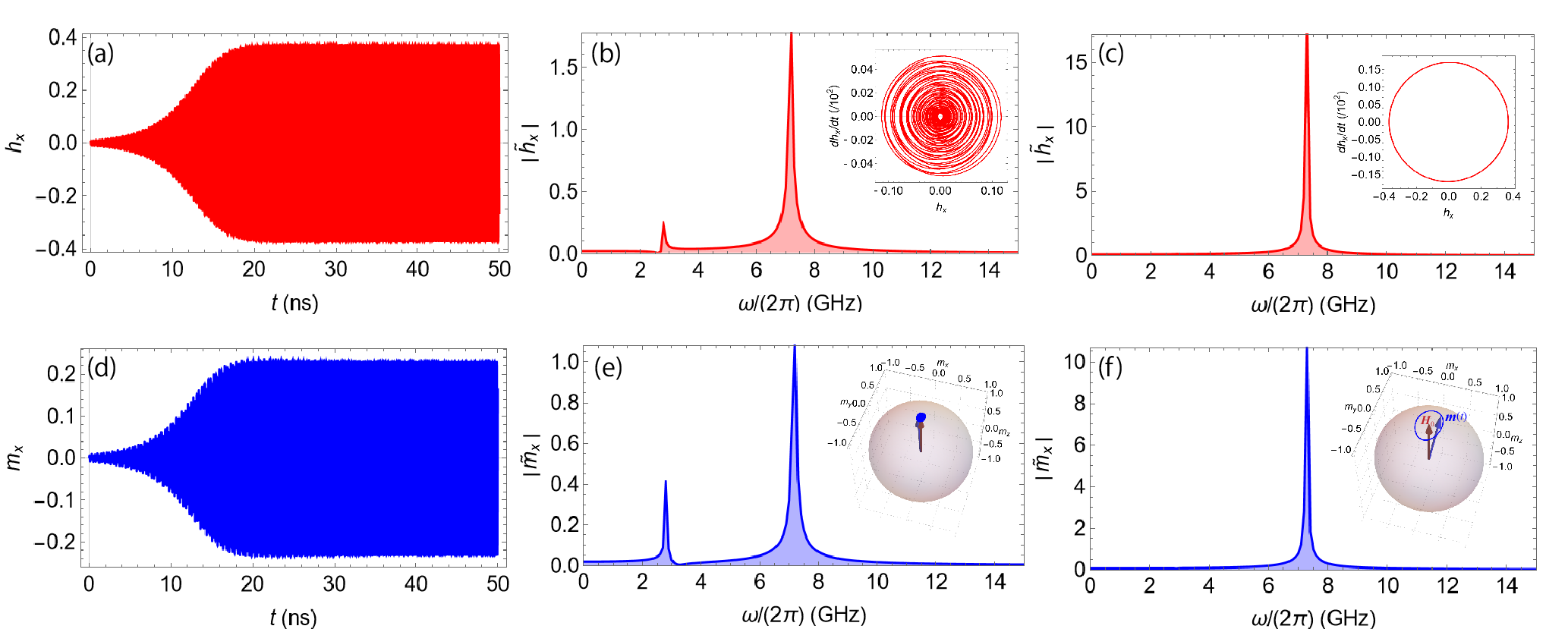} 
\par\end{centering}
\caption{Time evolution and its Fourier spectra of the gain-driven MP in the USC regime with $d_{\rm M}/d = 1$.
(a) Time evolution of the photon amplitude.
(b) Fourier spectrum of (a) in the time domain \(0 \leq t \leq 10\,\mathrm{ns}\).
(c) Fourier spectrum of (a) in the time domain \(40 \leq t \leq 50\,\mathrm{ns}\).
(d) Time evolution of the magnon amplitude.
(e) Fourier spectrum of (d) in the time domain \(0 \leq t \leq 10\,\mathrm{ns}\).
(f) Fourier spectrum of (d) in the time domain \(40 \leq t \leq 50\,\mathrm{ns}\).
Insets show the corresponding phase space trajectories of the photon and magnon.
The calculations are performed with the initial condition of $h_x(0) = 0.01$ and $m_x(0) = 0$ at the original modes crossing point ($\omega_{\rm m} = \omega_{\rm c}$).}
\label{Fig:USCGMP}
\end{figure*}

\subsection{Transient response}

\subsubsection{Strong coupling regime}

For the SC regime, Figs.~\ref{Fig:SCGMP}~(a) and (d) show the time evolution of photon ($h_x$) and magnon ($m_x$), respectively, wherein two distinct types of the oscillation pattern can be identified especially by Fig.~\ref{Fig:SCGMP}~(d). The first type appears in the early stage of the dynamics, roughly from $t=0$~ns to $t=10$~ns, and is characterized by rapid growth and transient behavior. The second type persists over a longer duration, at least from approximately $t=10$~ns to $t=50$~ns, indicating a nonlinearity that influences the entire course of the dynamics.

In order to elucidate the underlying cause of the two distinct oscillation patterns, let us discuss the corresponding Fourier spectrum of the oscillation, as shown in Figs.~\ref{Fig:SCGMP}~(b) and (c) for photons as well as Figs.~\ref{Fig:SCGMP}~(e) and (f) for magnons. Note that insets in each Fourier spectrum correspond to the phase space trajectories of photon and magnon.
At first, the oscillation pattern observed early (from $t=0$ to $10$~ns) in the dynamics is characterized by two major peaks seen in Fig.~\ref{Fig:SCGMP}~(b) or Fig.~\ref{Fig:SCGMP}~(e). These two peaks correspond to the energy splitting caused by the coherent coupling between photons and magnons. Therefore, the origin of the first type oscillation pattern is regarded as transient Rabi-like oscillations \cite{Gui24}. 
Secondly, the oscillation pattern observed after $t=10$~ns primarily reflects the generation of frequency-sidebands seen in Fig.~\ref{Fig:SCGMP}~(f), which attributes to the self-Kerr nonlinearity of magnons. In fact, the phase space trajectory of magnons in Fig.~\ref{Fig:SCGMP}~(f) exhibits a beating between $m_x~(m_y)$ and $m_z$ components, which is never observed within the linear dynamics \cite{Sugimoto20,Yemeli25}. 
However, this frequency-sideband generation is also transient. Until $t=500$~ns we have a stable auto-oscillation characterized by a single-mode peak at $4.42$~GHz. This is the main mode of magnons ($\omega_{\rm m}$) driven by a gain but its value visibly deviate from the auto-oscillation frequency $\Omega/(2\pi) = 4.80$~GHz anticipated by the analysis of the complex eigenfrequency [see Fig.~\ref{Fig:ReImomega}~(a)], which attributes to a frequency shift due to the self-Kerr nonlinearity. This issue will be addressed in greater detail subsequently.

\subsubsection{Ultrastrong coupling regime}

For the USC case, Figs.~\ref{Fig:USCGMP}~(a) and (d) show the time evolution of photon ($h_x$) and magnon ($m_x$), respectively. They appears to be more stable oscillation patterns than those of the SC case, as seen in Figs.~\ref{Fig:SCGMP}~(a) and (d). In fact, when we look at the corresponding Fourier spectra displayed in Figs.~\ref{Fig:USCGMP}~(b) and (c) for photons as well as Figs.~\ref{Fig:USCGMP}~(e) and (f) for magnons, we find a transient Rabi-like splitting at least in the duration from $t=0$~ns to $t=10$~ns, but in a flash we have a stable auto-oscillation characterized by a single-mode peak until $t=50$~ns. As shown in Figs.~\ref{Fig:USCGMP}~(c) and (f), the phase space trajectories of the photon and magnon exhibit limit cycles in each dynamics, respectively \cite{Tanekou23}.
Also, let us recall that in the SC case the onset time was determined to be approximately $500$~ns to achieve a stable auto-oscillation.
In addition, the main mode of magnons has $\omega_{\rm m}/(2\pi) = 7.28$~GHz, whose value is close to the auto-oscillation frequency $\Omega/(2\pi) = 7.16$~GHz anticipated by the analysis of the complex eigenfrequency [see Fig.~\ref{Fig:ReImomega}~(b)].
Therefore, we can mention that the onset time of the stable oscillation is shorter in the USC regime than in the SC regime, which attributes to the absence of the frequency-sideband generation in the USC regime. In other words, USC suppresses the frequency-sideband generation caused by the self-Kerr nonlinearity in the SC regime. 

\begin{figure}[ptb]
\begin{centering}
\includegraphics[width=0.45\textwidth,angle=0]{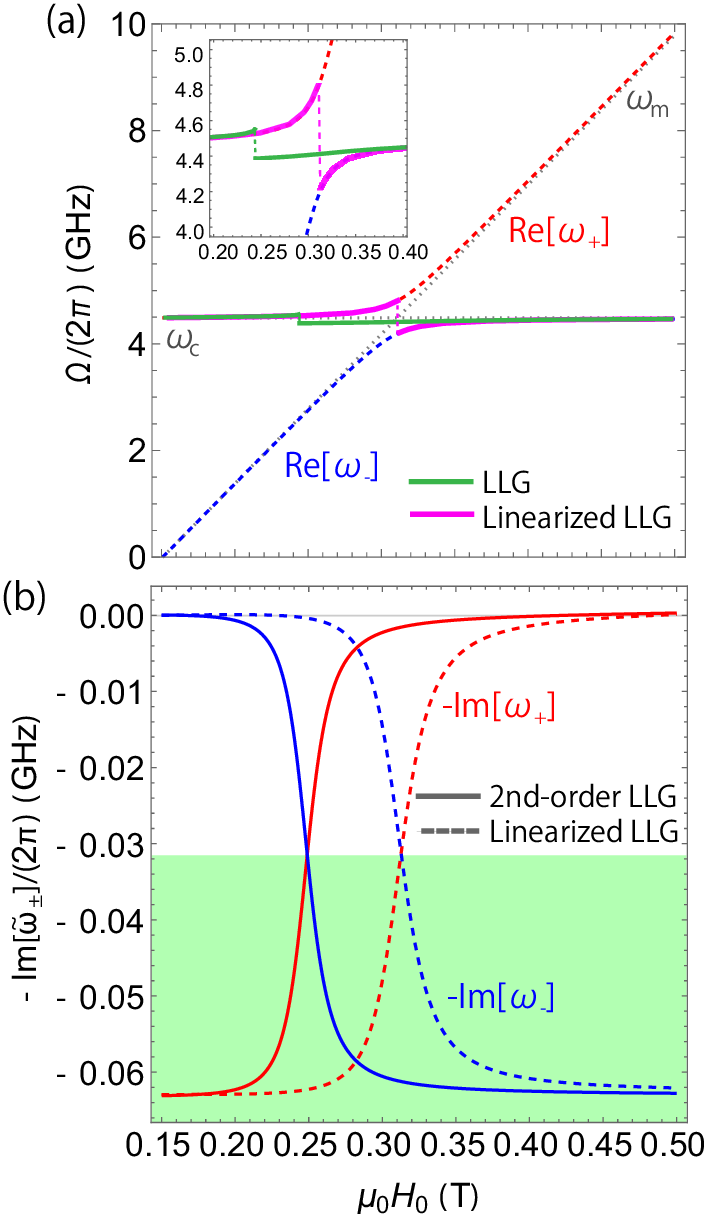} 
\par\end{centering}
\caption{(a) Auto-oscillation mode ($\Omega$) of the gain-driven MP for the SC case with $d_{\rm M}/d = 0.02$ as a function of an external magnetic field. The green (magenta) solid-line represents the solution obtained without (with) the linearized approximation in the LLG equation [Eq.~\eqref{LLGeq}]. Red and blue dashed liens ($\omega_\pm$) represent the real part of the complex eigenfrequency of the linearized gain-driven MP. Inset shows the zoom of the auto-oscillation mode. For calculating the auto-oscillation frequency, we used the magnetization dynamics between $t$ = 490~ns and 500~ns to obtain the corresponding Fourier spectrum. 
(b) Imaginary part of the complex eigenfrequencies with the second-order correction (the self-Kerr nonlinearity) as a function of an external magnetic field. 
}
\label{Fig:GMPmodeSC}
\end{figure}

\subsection{Auto-oscillation}

\subsubsection{Strong coupling regime}

Figure~\ref{Fig:GMPmodeSC}~(a) shows the auto-oscillation frequency $\Omega/(2\pi)$ of the gain-driven MP for the SC case with $d_{\rm M}/d = 0.02$.
The green solid-line represents the result obtained by using the LLG equation [Eq.~\eqref{LLGeq}] while the magenta solid-line represents the result obtained by using the linearized LLG equation [Eq.~\eqref{linearLLG}].
To clarify the impact of the nonlinear terms in the LLG equation [Eq.~\eqref{LLGeq}] on the auto-oscillation frequency, both frequencies are plotted together for comparison.  
Focusing on the green solid-line in Fig.~\ref{Fig:GMPmodeSC}~(a), we find that the point of the external magnetic field at which a maximum level repulsion occurs is shifted to the left side compared to the result of the linearized approximation (magenta solid-line). Note that each maximum level repulsion occurs at $\mu_0H_0 \approx 0.244$~T on the green solid-line and at $\mu_0H_0 \approx 0.311$~T on the magenta solid-line. This shift is identified with the frequency shift from $\Omega/(2\pi) = 4.80$~GHz to $4.42$~GHz previously  found in the transient response.
Furthermore, the magnitude of the level repulsion, which corresponds to the effective coupling strength $\tilde{g}$, is reduced compared to the result ($g$) obtained by the linearized LLG equation (magenta solid-line). In Fig.~\ref{Fig:GMPmodeSC}~(a), the value of $\tilde{g}$ at $\mu_0H_0 = 0.244$~T is given by $\tilde{g} \approx 0.24g$, which implies that the Kerr effect of magnons strongly suppresses the coherent magnon-photon coupling. Then, the gain-driven MP behaves like the original photon-mode driven by a gain, whose auto-oscillation frequency is mostly independent of the external magnetic field.

To begin our discussion on the nonlinear frequency shift and the suppression of the coupling strength in the SC regime, we use an approximated form of the LLG equation [Eq.~\eqref{2ndapproxLLG}]. 
Let us effectively include a correction of the second-order approximation [Eqs.~\eqref{Msm} and \eqref{alpham}] into the complex eigenfrequency of the linearized gain-driven MP.
At the steady-state of magnon dynamics, $m_{\perp}(t)$ in Eqs.~\eqref{Msm} and \eqref{alpham} is replaced by a constant value of $m_x$, defined by $m_{\infty} \equiv \lim_{t\to\infty}m_x(t)$. Namely, we have $\tilde{M}_{\rm s} = M_{\rm s}(1 - m_{\infty}^2/2)$ and $\tilde{\alpha} = \alpha(1 - m_{\infty}^2/2)$, by which we replace $M_{\rm s}$ and $\alpha$ in Eq.~\eqref{linearLLG} to take into account a correction of the second-order approximation. Then, magnon dynamics is governed by
\begin{align}
\frac{d^2 m_x}{d t^2}
+ 2\tilde{\alpha}\tilde{\omega}_{\rm m}\frac{d m_x}{d t}
+ \tilde{\omega}_{\rm m}^2m_x 
= \gamma\mu_0\tilde{M}_{\rm s}\tilde{\omega}_{\rm m}h_x
,\label{linearLLGnl}
\end{align}
where $\tilde{\omega}_{\rm m} = \omega_{\rm m} + \gamma \mu_{0}M_{\rm s}m_{\infty}^2/2$, $\tilde{M}_{\rm s} = M_{\rm s}(1 - m_{\infty}^2/2)$, and $\tilde{\alpha} = \alpha(1 - m_{\infty}^2/2)$.
In the same manner of Eq.~\eqref{LLG-RLC}, the corrected complex eigenfrequencies $\tilde{\omega}_\pm/(2\pi)$ are obtained by the determinant of 
\begin{align}
\bar{\Omega}_{\rm nl} = 
\begin{pmatrix}
\omega^2 + 2i(\beta - \beta_a)\omega_{\rm c}\omega - \omega_{\rm c}^2
& (d_{\rm M}/d)\omega^2 \\
\gamma\mu_0\tilde{M}_{\rm s}\tilde{\omega}_{\rm m}
& \omega^2
+ 2i\tilde{\alpha}\tilde{\omega}_{\rm m}\omega 
- \tilde{\omega}_{\rm m}^2
\end{pmatrix}
.\label{OmegaMPnl}
\end{align}
In Fig.~\ref{Fig:GMPmodeSC}~(b), we show the imaginary part of the complex eigenfrequencies as a function of the external magnetic field. For this calculation, we use $m_{\infty} = 0.914$ for $\mu_0H_0 = 0.244$~T at $t = 500$~ns obtained by solving Eqs.~\eqref{LLGeq} and \eqref{vdPeq}.
To clarify the impact of the second-order approximation on the imaginary part of the complex eigenfrequency, the result obtained by the linearized LLG is plotted together for comparison. For both cases, the green-shaded regions represent self-selection of bright modes, which is characterized by a sustained oscillation. Note that the dark modes that fade away within a limited time due to large dissipation do not appear. The switching point between two blight modes is the point at which the imaginary part of the complex eigenfrequencies crosses (damping exchange point).
As shown in Fig.~\ref{Fig:GMPmodeSC}~(b), the second-order correction, i.e., the self-Kerr nonlinearity, causes the damping exchange point to shift to the left side  (down to $\mu_0H_0 = 0.244$~T) from the original crossing point, which is good agreement with the shift in Fig.~\ref{Fig:GMPmodeSC}~(a). 
On the other hand, the coupling term between magnons and photons in the right hand side of Eq.~\eqref{linearLLGnl} indicates that the effective coupling strength $\tilde{g}$ is reduced by the self-Kerr nonlinearity via a modified salutation magnetization $\tilde{M}_{\rm s}$. By using $\tilde{M}_{\rm s}\approx0.58M_{\rm s}$ for $m_{\infty} = 0.914$, the estimated value is $\tilde{g} \approx 0.76g$. Let us recall that in Fig.~\ref{Fig:GMPmodeSC}~(a) the value of $\tilde{g}$ at $\mu_0H_0 = 0.244$~T is given by $\tilde{g} \approx 0.24g$. Hence, the effective coupling strength is overestimated for comparison with the result obtained by using Eq.~\eqref{LLGeq}, which may attribute to a higher-order correction of the approximation beyond Eq.~\eqref{2ndapproxLLG} and also to self-consistency of coupled nonlinear dynamics between magnon and photon subsystems.

\begin{figure}[ptb]
\begin{centering}
\includegraphics[width=0.45\textwidth,angle=0]{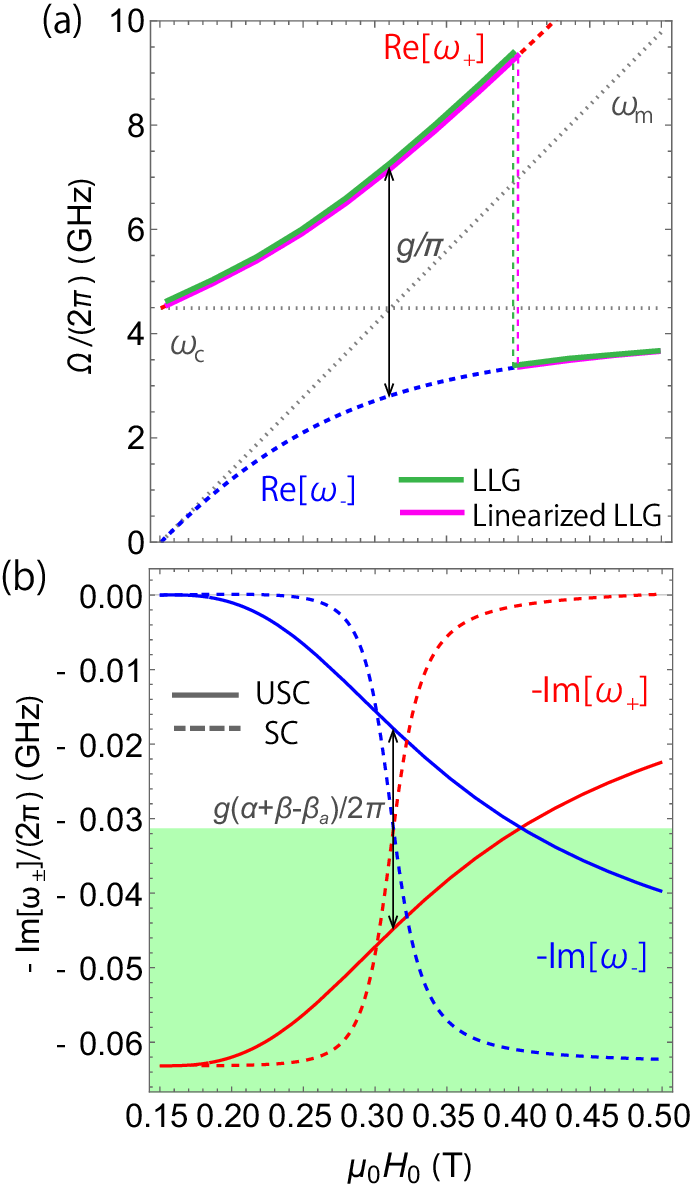} 
\par\end{centering}
\caption{(a) Auto-oscillation mode ($\Omega$) of the gain-driven MP for the USC case with $d_{\rm M}/d = 1$. For calculating the auto-oscillation frequency, we used the magnetization dynamics between $t$ = 490~ns and 500~ns to obtain the corresponding Fourier spectrum. 
(b) Imaginary part of the complex eigenfrequencies of the linearized gain-driven MP as a function of an external magnetic field. The solid-lines represent the SC case with $d_{\rm M}/d = 0.02$ and the dashed ones for the USC case with $d_{\rm M}/d = 1$.
}
\label{Fig:GMPmodeUSC}
\end{figure}

\subsubsection{Ultrastrong coupling regime}

Figure~\ref{Fig:GMPmodeUSC}~(a) shows the auto-oscillation frequency $\Omega/(2\pi)$ of the gain-driven MP for the USC case with $d_{\rm M}/d = 1$. The green and magenta solid-lines represent the results obtained without and with the linearized approximation in the LLG equation [Eq.~\eqref{LLGeq}], respectively. As seen, both lines appear to be almost same and to
strongly depend on the external magnetic field, indicating that the coherent magnon-photon coupling is much larger than the nonlinear interaction of magnons characterized by the self-Kerr nonlinearity. In other words, USC suppresses the frequency shift caused by the self-Kerr nonlinearity in the SC regime. 
This implies that nonlinear magnetization dynamics are suppressed under USC conditions and that there exists a trade-off relation between the coupling strength and the magnitude of the self-Kerr nonlinearity.

The trade-off relation can be intuitively understood by considering the behavior of the ferromagnet and the microwave magnetic field inside the inductor in each coupling regime (SC and USC).
In the SC regime, which has a small $d_{\rm M}/d$ (or $g/\omega_{\rm c}$) factor, the ferromagnet effectively behaves as an extremely thin film inserted into the inductor.
This results in a sufficiently large amplitude of magnetization dynamics relative to the microwave magnetic field inside the inductor, as seen in Figs.~\ref{Fig:SCGMP}~(a) and (d). Thus, a small $d_{\rm M}/d$ (or $g/\omega_{\rm c}$) factor leads to enhanced nonlinear magnetization dynamics.
In contrast, in the USC regime, which has a large $d_{\rm M}/d$ (or $g/\omega_{\rm c}$) factor, a thicker ferromagnet is inserted into the inductor compared to the SC case.
Substantially, the amplitude of magnetization dynamics remains small relative to the strength of the microwave magnetic field inside the inductor, as seen in Figs.~\ref{Fig:USCGMP}~(a) and (d). Therefore, a large $d_{\rm M}/d$ (or $g/\omega_{\rm c}$) factor contributes to the emergence of nearly linear magnetization dynamics.

Furthermore, in Fig.~\ref{Fig:GMPmodeUSC}~(a), we find that the point of the external magnetic field at which a maximum level repulsion occurs is shifted to the right side compared to the original modes crossing point at $\mu_0H_0\approx0.311$~T. Note that a maximum level repulsion occurs around $\mu_0H_0 \approx 0.397$~T on the green solid-line as well as the magenta solid-line.
To understand the origin of this shift, 
we analyze an imaginary part of the complex eigenfrequency of the linearized gain-driven MP based on Eq.~\eqref{OmegaMP} [see also Appendix~\ref{complex mode splitting}].
In Fig.~\ref{Fig:GMPmodeUSC}~(b), we show the imaginary part of the complex eigenfrequencies as a function of the external magnetic field. 
To clarify the impact of the USC on the imaginary part of the complex eigenfrequency, not only the USC case but also the SC case are plotted together for comparison. For both cases, the green-shaded regions represent self-selection of bright modes. As shown in Fig.~\ref{Fig:GMPmodeUSC}~(b), increasing the coupling strength causes the damping exchange point to shift to the right side (up to $\mu_0H_0 = 0.398$~T) from the original crossing point, which is good agreement with the shift in Fig.~\ref{Fig:GMPmodeUSC}~(a).

The origin of this shift can be understood by considering the complex mode splitting at the original modes crossing point [Eq.~\eqref{Shift}].
Equation~\eqref{Shift} indicates that a gap opens in the imaginary parts of the complex eigenfrequencies at the original modes crossing point, resulting in the observed auto-oscillation frequency shift in Fig.~\ref{Fig:GMPmodeUSC}~(a). Hence, the origin of the auto-oscillation frequency shift is related with the magnon-photon coupling and gain.
Based on Eq.~\eqref{Shift}, in the SC regime with a relatively small coupling strength, the shift of the damping exchange point appears to be absent. In  Fig.~\ref{Fig:GMPmodeUSC}~(a), we find that the large shift of the damping exchange point in the USC case results in the emergence of the magnon-like mode in $0.3~\mathrm{T} \le \mu_{0}H_{0} \le 0.4~\mathrm{T}$ compared to the SC case. Since this feature is prominent in the USC regime, suggesting that, for applications such as maser-like devices, the USC is advantageous owing to widely tunable frequency range and controllability of auto-oscillation frequency by means of the external magnetic field.

\section{Conclusion}
\label{Conclusion}

In conclusion, we have theoretically investigated the nonlinear dynamics of gain-driven MPs based on an effective circuit model that combines the $RLC$-circuit equation for photons and the LLG equation for magnons. 
Numerical calculations revealed that in the SC regime magnonic self-Kerr nonlinearity plays a dominant role in the dynamics of magnon, resulting in transient frequency-sidebands generation and a frequency shift. 
These nonlinear effects suppress the coherent magnon-photon coupling in the gain-driven MP. 
In contrast, in the USC regime, the self-Kerr nonlinearity is effectively reduced, leading to stable and nearly linear oscillations of the gain-driven MP. 
Consequently, the auto-oscillation mode exhibits magnon-like features over a broader range of external magnetic fields. 

Theoretical analysis based on the LLG equation with suitable approximations confirmed a trade-off relation between the coupling strength and the self-Kerr nonlinearity: increasing the coupling strength suppresses the observed nonlinear behavior, thereby enhancing stability and tunability of the gain-driven MP. 
Furthermore, the USC induced a frequency shift in the auto-osculation mode due to a shift of the crossing point of the imaginary part of the complex eigenfrequency, which is attributed to the interplay between gain-loss and USC in polariton dynamics.
These results suggest that ultrastrongly coupled MP systems provide a promising platform for developing maser-like devices \cite{Hou21,Yao23} with widely tunable frequency range and controllability by means of the external magnetic field.
In addition, the present study suggests that magnon (spin) auto-oscillation can be realized by utilizing gain-driven hybrid systems, i.e., gain-driven coherent magnetization dynamics, which may provide a novel operational principle for spintronic oscillators such as spin-torque oscillators \cite{Hou21,Jiang24}.

\begin{acknowledgments}
The authors thank G. E. W. Bauer, T. Hioki, T. Yamashita, T. Otaki, S. Tomita, and T. Taniguchi for valuable discussions. This work was supported by Grants-in-Aid for Scientific research (Grants No.~22K14591, No.~24K00563, and No.~25H02105). R. S. thanks to GP-Spin program at Tohoku University. H. M. acknowledges support from CSIS at Tohoku University.
\end{acknowledgments} 

\section*{Data Availability}

The data that support the findings of this article are not publicly available upon publication because it is not technically feasible and/or the cost of preparing, depositing, and hosting the data would be prohibitive within the terms of this research project. The data are available from the authors upon reasonable request.

\appendix

\section{Derivation of complex mode splitting}
\label{complex mode splitting}

\begin{figure}[ptb]
\begin{centering}
\includegraphics[width=0.44\textwidth,angle=0]{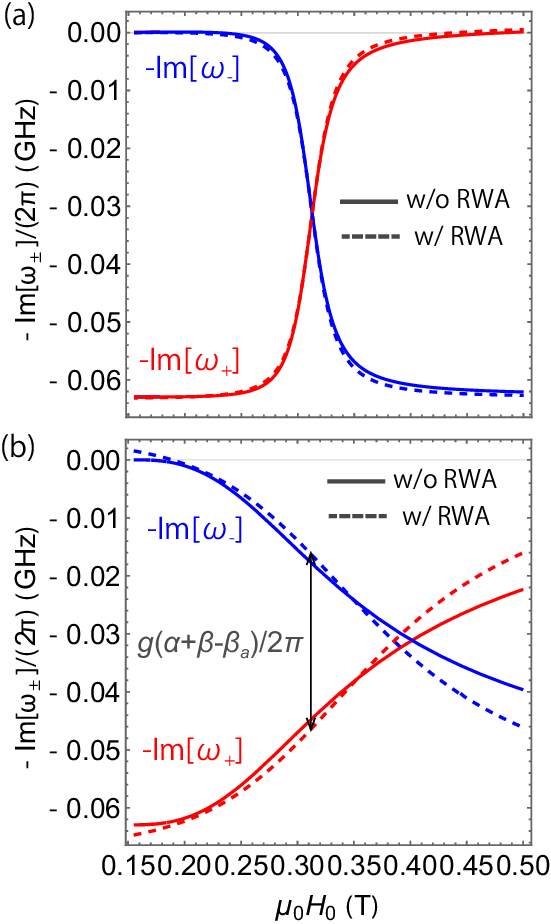} 
\par\end{centering}
\caption{
Imaginary part of the complex eigenfrequencies of the linearized gain-driven MP as a function of an external magnetic field. (a) A SC case with $d_{\rm M}/d = 0.02$ and (b) USC case with $d_{\rm M}/d = 1$. 
}
\label{Fig:ImaginaryShift}
\end{figure}

The complex eigenfrequencies $\omega_\pm/(2\pi)$ of the linearized gain-driven MP are obtained by solving the determinant of Eq.~\eqref{OmegaMP}:
\begin{align}
&\left(  \omega^2 + 2i\left(  \beta - \beta_a\right)\omega_{\rm c}\omega - \omega_{\rm c}^2\right)\left(  \omega^2
+ 2i\alpha\omega_{\rm m}\omega - \omega_{\rm m}^2\right)\nonumber\\
 &- (d_{\rm M}/d)\gamma\mu_0M_{\rm s}\omega_{\rm m}\omega^2 = 0
.\label{detOmegaMP}
\end{align}
At the original modes crossing point, let us introduce a classical version of the RWA \cite{Chiba24APL,Chiba24MSJ,Chiba25}, namely, $\omega^2 - \omega_{\rm c}^2 \approx 2\omega_{\rm c}\left(  \omega - \omega_{\rm c}\right)$ and $\omega^2 - \omega_{\rm m}^2 \approx 2\omega_{\rm c}\left(  \omega - \omega_{\rm m}\right)$, to Eq.~(\ref{detOmegaMP}). Furthermore, writing $\omega = \omega_{\rm c} + \delta$ with $\delta \ll \omega$, we neglect the terms of order $\delta^2$ and $\delta(d_{\rm M}/d)\gamma\mu_0M_{\rm s}$, leading to 
\begin{align}
\left(  \omega + i\left(  \beta - \beta_a\right)\omega - \omega_{\rm c}\right)\left(  \omega
+ i\alpha\omega - \omega_{\rm m}\right)- g^2 = 0
,\label{detOmegaMPRWA}
\end{align}
where $g = \sqrt{(d_{\rm M}/d)\gamma\mu_0M_{\rm s}\omega_{\rm c}}/2$. Then, we find the complex eigenfrequencies at the original modes crossing point
\begin{align}
\omega_{\pm} = \omega_{\rm c} \pm g - \frac{1}{2}\left(  \alpha + \beta - \beta_a\right)\left(  \omega_{\rm c} \pm g\right)i
.\label{}
\end{align}
Therefore, the complex mode splitting is given by Eq.~\eqref{Shift}.

Figure~\ref{Fig:ImaginaryShift} shows the imaginary part of the complex eigenfrequencies of the linearized gain-driven MP with (without) RWA in each coupling regime. As shown in Fig.~\ref{Fig:ImaginaryShift}~(a), in the SC regime, the RWA appears to be a good approximation which well reproduces the result of the w/o RWA. In contrast, in the USC regime, one can find a visible deviation of the damping exchange point between the two calculation cases in Fig.~\ref{Fig:ImaginaryShift}~(b), indicating that a nontrivial frequency shift such as BS shift affects the self-selection of the gain-driven MP.


\begin{thebibliography}{999}

\bibitem{Harder21}M. Harder, B. M. Yao, Y. S. Gui, C.-M. Hu, Coherent and dissipative cavity magnonics, J. Appl. Phys. \textbf{129}, 201101 (2021).

\bibitem{Rameshti22}B. Z. Rameshti, S. V. Kusminskiy, J. A. Haigh, K. Usami, D. L.-Quirion, Y. Nakamura, C.-M. Hu, H. X. Tang, G. E. Bauer, and Y. M. Blanter, Cavity magnonics, Phys. Rep. {\bf 979}, 1 (2022).

\bibitem{Yuan22}H.Y. Yuan, Y. Cao, A. Kamra, R. A. Duine, P. Yan, Quantum magnonics: when magnon spintronics meets quantum information science, Phys. Rep. {\bf 965}, 26 (2022).

\bibitem{Quirion20}D. L.-Quirion, S. P. Wolski, Y. Tabuchi, S. Kono, K. Usami, and Y. Nakamura, 
Entanglement-based single-shot detection of a single magnon with a superconducting qubit, Science \textbf{367}, 425 (2020). 

\bibitem{Zhang15}X. Zhang, C.-L. Zou, N. Zhu, F. Marquardt, L. Jiang, and H. X. Tang, Magnon dark modes and gradient memory, Nat. Commun. {\bf 6}, 8914 (2015).

\bibitem{Shen21}R.-C. Shen, Y.-P. Wang, J. Li, S.-Y. Zhu, G. S. Agarwal, and J. Q. You, Long-Time Memory and Ternary Logic Gate Using a Multistable Cavity Magnonic System, Phys. Rev. Lett. \textbf{127}, 183202 (2021). 

\bibitem{Hou21}J. T. Hou, P. Zhang, and L. Liu, Proposal for a Spin-Torque-Oscillator Maser Enabled by Microwave Photon-Spin
Coupling, Phys. Rev. Appl. \textbf{16}, 034034 (2021).

\bibitem{Yao23}B. Yao, Y. S. Gui, J. W. Rao, Y. H. Zhang, W. Lu, and C.-M. Hu, Coherent microwave emission of gain-driven polaritons, Phys. Rev. Lett. {\bf 130}, 146702 (2023).

\bibitem{Diaz19}P. F.-D\'{i}az, L. Lamata, E. Rico, J. Kono, and E. Solano, Ultrastrong coupling regimes of light-matter interaction, Rev. Mod. Phys. {\bf 91} 025005 (2019).

\bibitem{Kockum19}A. F. Kockum, A. Miranowicz, S. D. Liberato, S. Savasta, and F. Nori, Ultrastrong coupling between light and matter, Nat. Rev. Phys. {\bf 1}, 19 (2019).

\bibitem{Baydin25}A. Baydin, H. Zhu, M. Bamba, K. R. A. Hazzard, and J. Kono, Perspective on the quantum vacuum in matter, Opt. Mater. Express {\bf 8}, 1833 (2025).

\bibitem{Diaz10}P. F.-D\'{i}az, J. Lisenfeld, D. Marcos, J. J. Garc\'{i}a-Ripoll, E. Solano, C. J. P. M. Harmans, and J. E. Mooij, Observation of the Bloch-Siegert shift in a qubit-oscillator system in the ultrastrong coupling regime, Phys. Rev. Lett. {\bf 105}, 237001 (2010).

\bibitem{Li18}X. Li, M. Bamba, Q. Zhang, S. Fallahi, G. C. Gardner, W. Gao, M. Lou, K. Yoshioka, M. J. Manfra, and J. Kono, Vacuum Bloch-Siegert Shift in Landau Polaritons with Ultra-high Cooperativity, Nat. Photonics {\bf 12}, 324 (2018).

\bibitem{Golovchanskiy21SciAdv}I. A. Golovchanskiy, N. N. Abramov, V. S. Stolyarov, M. Weides, V. V. Ryazanov, A. A. Golubov, A. V. Ustinov, and M. Y. Kupriyanov, Ultrastrong photon-to-magnon coupling in multilayered heterostructures involving superconducting coherence via ferromagnetic layers, Sci. Adv. {\bf 7}, eabe8638 (2021).

\bibitem{Golovchanskiy21PRAP}I. Golovchanskiy, N. N. Abramov, V. S. Stolyarov, A. A. Golubov, M. Yu. Kupriyanov, V. V. Ryazanov, and A. V. Ustinov, Approaching deep-strong on-chip photon-tomagnon coupling, Phys. Rev. Appl. {\bf 16}, 034029 (2021).

\bibitem{Ghirri23}A. Ghirri, C. Bonizzoni, M. Maksutoglu, A. Mercurio, O. Di Stefano, S. Savasta, and M. Affronte, Ultrastrong magnon-photon coupling achieved by magnetic films in contact with superconducting resonators, Phys. Rev. Appl. {\bf 20}, 024039 (2023).

\bibitem{Bourcin23}G. Bourcin, J. Bourhill, V. Vlaminck, and V. Castel, Strong to ultrastrong coherent coupling measurements in a YIG/cavity system at room temperature, Phys. Rev. B {\bf 107}, 214423 (2023).

\bibitem{Mita25PRAP}K. Mita, T. Chiba, T. Kodama, T. Ueda, T. Nakanishi, K. Sawada, S. Tomita, Ultrastrongly coupled and directionally nonreciprocal magnon polaritons in magnetochiral metamolecules, Phys. Rev. Appl. {\bf 23}, L011004 (2025).

\bibitem{Mita25arXiv}K. Mita, T. Kodama, T. Nakanishi, T. Ueda, K. Sawada, T. Chiba, S. Tomita, Microwave One-way Transparency by Large Synthetic Motion of Magnetochiral Polaritons in Metamolecules,
arXiv:2503.22279.


\bibitem{Zhang25}C. Zhang, M. Kim, Y.-H. Zhang, Y.-P. Wang, D. Trivedi, A. Krasnok, J. Wang, D. Isleifson, R. Roshko, and C.-M. Hu, Gainloss coupled systems, APL Quantum {\bf 2}, 011501 (2025).

\bibitem{Kim24}M. Kim, C.Zhang, C. Lu, C.-M. Hu, Low phase noise microwave oscillator based on gain driven polariton, Appl. Phys. Lett. \textbf{124}, 114103 (2024).

\bibitem{Gui24}Y. S. Gui and C.-M. Hu, Transient response of a gain-driven polariton, Phys. Rev. Appl. \textbf{21}, 044023 (2024).

\bibitem{Lu24}C. Lu, M. Kim, C. Zhang, C.-M. Hu, Solid-state injection locking microwave amplifier, J. Appl. Phys. \textbf{136}, 053903 (2024).

\bibitem{Gui25}Y. S. Gui, G. S. Kozyniak, and C.-M. Hu, Broadband microwave emission from acoustically
modulated gain-driven polaritons, J. Appl. Phys. \textbf{137}, 043905 (2025).

\bibitem{Rao23}J. Rao, C. Y. Wang, B. Yao, Z. J. Chen, K. X. Zhao, and W. Lu, Meterscale Strong Coupling between Magnons and Photons, Phys. Rev. Lett. {\bf 131}, 106702 (2023).

\bibitem{Zheng23}S. Zheng, Z. Wang, Y. Wang, F. Sun, Q. He, P. Yan, and H. Y. Yuan, Tutorial: Nonlinear magnonics, J. Appl. Phys. {\bf 134}, 151101 (2023).

\bibitem{Elyasi20}M. Elyasi, Y. M. Blanter, and G. E. W. Bauer, Resources of nonlinear cavity magnonics for quantum information, Phys. Rev. B {\bf 101}, 054402 (2020).

\bibitem{Tatsumi25}R. Tatsumi, T. Chiba, T. Komine, and H. Matsueda, Chaotic magnetization dynamics in magnetic Duffing oscillator, Phys. Rev. E {\bf 111}, 064202 (2025).

\bibitem{Wang16}Y.-P. Wang, G.-Q. Zhang, D. Zhang, X.-Q. Luo, W. Xiong, S.-P.Wang, T.-F. Li, C.-M. Hu, and J. Q. You, Magnon Kerr effect in a strongly coupled cavity-magnon system, Phys. Rev. B {\bf 94}, 224410 (2016).

\bibitem{Wang18}Y.-P.Wang, G.-Q. Zhang, D. Zhang, T.-F. Li, C.-M. Hu, and J. Q. You, Bistability of cavity magnon polaritons, Phys. Rev. Lett. {\bf 120}, 057202 (2018).

\bibitem{Bai15}L. Bai, M. Harder, Y. P. Chen, X. Fan, J. Q. Xiao, and C.-M. Hu, Spin Pumping in Electrodynamically Coupled Magnon-Photon Systems, Phys. Rev. Lett. {\bf 114}, 227201 (2015).

\bibitem{Grigoryan18}V. L. Grigoryan, K. Shen, and K. Xia, Synchronized spin-photon coupling in a microwave cavity, Phys. Rev. B {\bf 98}, 024406 (2018).

\bibitem{Chiba24APL}T. Chiba, T. Komine, and T. Aono, Ultrastrong-coupled magnon-polariton in a dynamical inductor based on magnetic-insulator/topological-insulator bilayers, Appl. Phys. Lett. {\bf 124}, 012402 (2024).

\bibitem{Chiba24MSJ}T. Chiba, T. Komine, and T. Aono, Microwave Transmission Theory for On-Chip Ultrastrong-Coupled Magnon-Polariton in Dynamical Inductors, J. Mag. Soc. Jpn. {\bf 48}, 21 (2024).

\bibitem{Chiba25}T. Chiba, R. Suzuki, T. Otaki, and H. Matsueda, Circuit-based cavity magnonics in the ultrasrtong and deep-strong coupling regimes, submitted.

\bibitem{Schreier15}M. Schreier, T. Chiba, A. Niedermayr, J. Lotze, H. Huebl, S. Gep\''{r}ags, S. Takahashi, G. E. W. Bauer, R. Gross, and S. T. B. Goennenwein, Phys. Rev. B {\bf 92}, 144411 (2015).

\bibitem{Zhang24van}C. Zhang , M. Kim, J. Wang, and C.-M. Hu, Van der Pol-Duffing oscillator and its application to gain-driven light-matter interaction, 
Phys. Rev. Appl. \textbf{22}, 014034 (2024).

\bibitem{Tanekou23}S. T. Tanekou, J. Ramadoss, J. Kengne, G. D. Kenmoe, and K. Rajagopal, Coexistence of Periodic, Chaotic and Hyperchaotic Attractors in a System Consisting of a Duffing Oscillator Coupled to a van der Pol Oscillator, Int. J. Bifurcation and Chaos {\bf 33}, 2330004 (2023).

\bibitem{Cheng16}R. Cheng, D. Xiao, and A. Brataas, Terahertz antiferromagnetic spin Hall nano-oscillator, Phys. Rev. Lett. {\bf 116}, 207603 (2016).

\bibitem{Sugimoto20}S. Sugimoto, S. Iwakiri, Y. Kozuka, Y. Takahashi, Y. Niimi, K. Kobayashi, S. Kasai, Multiple modes of a single spin torque oscillator under the non-linear region, AIP Adv. {\bf 10}, 075115 (2020).

\bibitem{Yemeli25}I. N. Yemeli, S. Perna, D. Gou\'{e}r\'{e}, A. Kolli, S. Sangiao, J. M. De Teresa, M. Mu\~{n}oz, A. Anane, M. d'Aquino \textit{et al.}, Self-Modulation Instability in High Power Ferromagnetic Resonance of BiYIG Nanodisks, Phys. Rev. Lett. {\bf 135}, 056703 (2025).

\bibitem{Jiang24}S. Jiang, L. Yao, S. Wang, D. Wang, L. Liu, A. Kumar, A. A. Awad, A. Litvinenko, M. Ahlberg, R. Khymyn, S. Chung, G. Xing, and J. \AA kerman, Spin-torque nano-oscillators and their applications, Appl. Phys. Rev. {\bf 11}, 041309 (2024).

\end{thebibliography}

\end{document}